\documentclass{article}

\usepackage[T1]{fontenc}
\usepackage[utf8]{inputenc}
\usepackage[sort]{cite}
\usepackage{siunitx}
\usepackage{color}
\usepackage{multirow}
\usepackage{booktabs}
\usepackage{tabularx}
\usepackage{acronym}
\usepackage{makecell}
\usepackage{spconf,amsmath,graphicx}
\usepackage[hidelinks]{hyperref}
\usepackage{url}

\newcommand{\Figure}[1]{Figure~\ref{#1}}
\newcommand{\Table}[1]{Table~\ref{#1}}

\acrodef{DNN}[DNN]{Deep Neural Network}
\acrodef{DE}[DE]{Dialogue Enhancement}
\acrodef{qPSM}[qPSM]{quantized Phase-Sensitive Mask}
\acrodef{PSM}[PSM]{Phase-Sensitive Mask}
\acrodef{qms}[qms]{quantized mid-side}

\title{ODAQ-Collection: Expanding the open dataset of audio quality \\ with a focus on spatial distortions}

\name{\begin{tabular}{c}Chih-Wei~Wu$^{1*}$, Matteo~Torcoli$^{2*}$, Mhd~Modar~Halimeh$^{3}$, Pablo~M.~Delgado$^{4}$, Sascha~Dick$^{4}$, \\
      Christoph~Thompson$^{5}$, Phillip~A.~Williams$^{1}$, Emanu\"el~A.~P.~Habets$^{4}$\end{tabular}\thanks{*Co-first authors.}}
\address{$^{1}$\small{Netflix, Inc., Los Gatos, USA} \\
         $^{2}$\small{Amplifon S.p.A., Milan, Italy} \\
         $^{3}$\small{Starkey Hearing Technologies, Minnesota, USA} \\
         $^{4}$\small{Fraunhofer Institute for Integrated Circuits IIS, Erlangen, Germany} \\
         $^{5}$\small{Ball State University, Muncie, USA}}

\begin{document}
\maketitle

\begin{abstract}
The Open Dataset of Audio Quality (ODAQ) is an expanding resource that addresses the scarcity of openly available data in audio quality assessment research.
While previous contributions focused mainly on timbral distortions, with spatial processing limited to basic mid-side and left-right stereo conditions, we present a new contribution featuring high-quality music and speech recordings with diverse stereo images, controlled spatial distortions, and subjective listening test results from trained listeners in a laboratory setting.
With 148 pairs of stereo audio samples processed by three different method classes, two applying controlled binaural cue distortions and one applying dialogue enhancement, this addition expands the collection to 688 pairs of processed and unprocessed audio signals spanning 17 processing method classes, along with 18,984 raw subjective ratings from listeners.
All associated data and code are released as open resources to facilitate future expansions and research on objective audio quality metrics.
\end{abstract}


\section{Introduction}\label{sec:introduction}

Quantifying perceived audio quality is essential for enabling measurable progress across various audio tasks such as source separation\cite{emiya2011subjective}, audio coding\cite{erne2001perceptual}, and music generation\cite{vinay2022evaluating}.
Although subjective listening tests remain the gold standard for audio quality assessment, objective audio quality metrics are often considered a convenient alternative due to their practicality and compatibility across different application domains \cite{torcoli2021objective}.
Developing such metrics, however, requires well-curated datasets to characterize human perception of audio quality, inform the design of a computational approach, and validate its correlation with human ratings.
Creating these datasets is time-consuming and costly, and the sharing of audio source files can sometimes be prohibited due to copyright restrictions. As a result, many datasets for this purpose remain private\cite{dick2018generation} or are only shared to a limited extent\cite{ISO2011USAC}.
While individual datasets exist for specific application domains such as audio coding\cite{PublicMultiformatTest2014} and source separation\cite{emiya2011subjective}, they are typically one-off contributions limited in size and scope of experimental conditions.
Larger datasets have since been constructed via crowdsourcing\cite{cartwright2018crowdsourced, mittag2021nisqa, maniati2022somos} or by outsourcing evaluations to external vendors\cite{tjandra2025meta}, but these often involve uncontrolled listening environments and increased label noise, necessitating sophisticated pre- and post-screening procedures\cite{lechler2025crowdsourcing}.

To address these limitations, the Open Dataset of Audio Quality (ODAQ) \cite{Torcoli2024ODAQ} was proposed as a publicly available resource featuring high-quality audio signals and subjective scores collected from the Multiple Stimuli with Hidden Reference and Anchor (MUSHRA)\cite{MUSHRA2015} tests conducted in controlled listening environments.
After the initial release, a subsequent addition~\cite{Dick2024ODAQBSU} provided further subjective scores based on the same material and protocol, confirming the reproducibility of the initial test. This addition also introduced a basic benchmark of objective quality metrics on the dataset.
A later addition introduced new audio content investigating basic stereo processing conditions, namely mid-side (MS) and left-right (LR), together with MUSHRA test results allowing comparisons between preprocessing methods and presentation context\cite{Dick2025investigating}, and a companion study benchmarked objective quality metrics on this material\cite{delgado2025exploring}.
A more recent addition from the community \cite{lanzendorfer2025evaluating} added derived assets from the initial release processed by neural audio codecs. In contrast to the other additions, subjective ratings were collected via crowdsourced MUSHRA tests with post-screening, and the participants were self-reported expert listeners.

\begin{table}[t]
\footnotesize
\centering
\caption{Processing methods and quality levels (Q6 is the highest quality) in the proposed addition.}
\begin{tabular*}{\columnwidth}{@{\extracolsep{\fill}}cccc}
\toprule
                                & \textbf{ICC}& \textbf{ILD}& \textbf{DE}            \\
                                & {\footnotesize $d$} & {\footnotesize[\# quant. levels]} & {\footnotesize (side bits, mid bits)}            \\
\midrule
Q1                              & Overflow          & 0\,dB (fixed)        & DeepFilterNet\cite{schroter2022deepfilternet} \\
Q2                              & Collapse         & 5          & Bandit-v2\cite{watcharasupat2024}     \\
Q3                              & 0.99        & 7          & qmsPSM (4, 2) \\
Q4                              & 0.0         & 100        & qmsPSM (3, 3) \\
Q5                              & -           & -          & qmsPSM (3, 5) \\
Q6                              & -           & -          & qPSM          \\
\bottomrule
\end{tabular*}
\vspace{-1em}
\label{tab:quallevels}
\end{table}

While the scope has expanded over these contributions, earlier additions have focused mainly on timbral quality, with diverse spatial distortions largely absent from the existing audio content.
In addition, as contributions accumulate, the resource becomes increasingly fragmented, making it less straightforward for prospective contributors to see the overall picture and identify open problems, and for end users to aggregate results across subsets coherently.
To fill these gaps and support the continued evolution of this open resource, this work introduces a new subset focused on stereo spatial distortions, consisting of high-quality speech and music stereo recordings (Sec.~\ref{subsec:audio-material}) processed under controlled binaural cue distortions (Sec.~\ref{subsec:processing1}) and dialogue enhancement (Sec.~\ref{subsec:processing2}), along with the corresponding MUSHRA results (Sec.~\ref{subsubsec:listening}). It also organizes the overall collection into a unified resource (Sec.~\ref{sec:overview}) with scripts for aggregating results across subsets.

\vspace{-1.0em}

\section{Proposed Addition}\label{sec:dataset} \vspace{-0.5em}

\subsection{Audio Material}\label{subsec:audio-material} \vspace{-0.5em}

All audio source material in our proposed addition consists of original recordings made in a musical performance hall with an average RT60 of approximately \SI{2.5}{\second}, with the exception of one pop music composition.
In preparation, calibrated test recordings were made to determine the ideal direct sound to diffuse field ratios for the main microphone array, and a distance of 1.5 to \SI{2}{\metre} was selected.
All sources were recorded at \SI{96}{\kilo\hertz} and 24~bits, then mixed to stereo and exported at \SI{48}{\kilo\hertz} and 24~bits.

The proposed addition contains 18 unique audio sources of speech and music content.
The speech sources include dialogue-only examples as well as mixtures of dialogue and music, balanced between male and female speakers;
two unique sources feature animated dialogue between a male and a female speaker. Voice acting emphasizes emotional and expressive delivery, aiming to mimic high-quality cinematic content and challenge source separation systems.
The music sources include monophonic solo instruments and polyphonic mixtures, ranging from a few isolated notes to an excerpt of classical music performed by a string quartet and an excerpt from a pop composition.
The stereo images across these sources are intentionally diverse, including artificially panned moving sources, hard-panned left/right placement, and centered sources with a wide, naturally diffuse field captured during recording. The resulting subset, including processed items and anchor conditions, contains 148 pairs of audio signals with an average duration of \SI{11.8}{\second}.

All sources were processed with three methods representing distinct quality levels; the configuration is summarized in \Table{tab:quallevels}, with details in the sections below.

\vspace{-1.0em}

\subsection{Binaural Cue Processing}\label{subsec:processing1} \vspace{-0.5em}

\begin{figure}[t]
    \centering
    \includegraphics[width=\columnwidth, trim=0.8cm 0 0.75cm 0, clip]{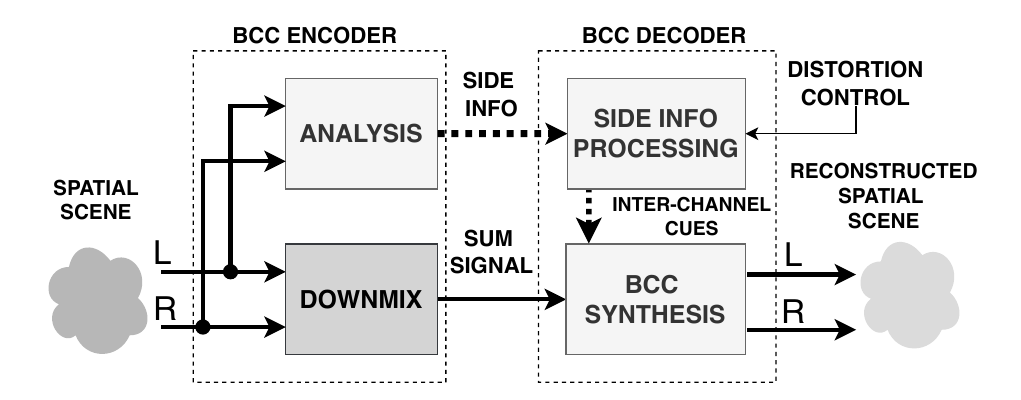}
    \caption{Generation of spatial distortions using a modified Binaural Cue Codec \cite{faller2003binaural} following \cite{delgado2019investigations}.} \vspace{-1em}
    \label{fig:BCC_codec}
\end{figure}

We modified a Binaural Cue Coding (BCC) perceptual audio codec \cite{faller2003binaural} to distort the stereo field. BCC represents the technology underlying modern stereo and spatial audio codecs, which quantize and transmit parametric inter-channel cues to preserve the stereo image at low bitrates \cite{breebaart2005parametric}.
As shown in \Figure{fig:BCC_codec}, manipulating this side information introduces isolated Interaural Level Differences (ILD) and Interaural Cross-Correlation (ICC) artifacts, causing perceived deviations in sound source localization and reduced Apparent Source Width (ASW) \cite{blauert1986spatial} relative to the unprocessed signal.
The BCC uses $21$~ms analysis windows with $10.7$~ms overlap at $48$~kHz sample rate to generate per-frame ICC and ILD distortions.
We assume headphone presentation of the signals, meaning that modifications in inter-channel cues correspond to similar interaural cue differences \cite{faller2003binaural}.

\textbf{Generation of ILD Distortions}:
The BCC codec reconstructs the stereo signal by applying per-band inter-channel level difference across Equivalent Rectangular Bandwidth (ERB)-spaced frequency bands \cite{moore1983suggested}.
In its high-quality configuration (Q4), the ILD span of $\pm$24\,dB is uniformly quantized into 100 levels (step size $\Delta = 0.48$\,dB), falling below the Just Noticeable Difference (JND) of 0.5\,dB for ILD \cite{breebaart2005parametric}.
At Q3 and Q2, the quantization levels are reduced to 7 and 5, respectively, producing clearly audible lateral jumps in source position.
Finally, Q1 represents the maximum ILD distortion condition by setting all per-band inter-channel level differences to 0\,dB, collapsing the entire ILD-generated stereo span, and causing all auditory sources to be perceived at the center of the image. The ICC reconstruction path of the BCC remains unmodified.

\textbf{Generation of ICC Distortions}:
The BCC codec estimates per-band ICC as a normalized cross-correlation coefficient for each ERB-spaced frequency band.
The ICC level quantizer computes a distorted ICC value at a specified perceptual distance from the reference, based on the psychometric model of ICC sensitivity reported by~\cite{culling2001interaural}.
In this model, the sensitivity index can be interpreted as a perceptual distance $d$ between two ICC values, with $|d|=1$ corresponding to one JND.
The model relates $d$ to the reference and distorted ICC (denoted by $\text{ICC}_\text{ref}$ and $\text{ICC}_\text{dist}$, respectively) via frequency-dependent functions $k(f)$ and $g(f)$:
\begin{equation}
   \label{eq:iacc_sensitivity}
    d = e^{k(f) \cdot \text{ICC}_\text{dist} + g(f)} -
    e^{k(f) \cdot \text{ICC}_\text{ref} + g(f)},
\end{equation}
where $k(f)$ and $g(f)$ are sigmoids centered around 600--700\,Hz with parameters defined as in \cite{culling2001interaural}, emphasizing ICC changes at low frequencies, to which humans are most sensitive.
Based on Eq.~\eqref{eq:iacc_sensitivity}, we propose to generate signals with distorted ICC values $\text{ICC}_\text{gen}$ applied at each ERB band $m$ with center frequency $f_m$:
\begin{equation}
    \text{ICC}_\text{gen}(m) =
    \frac{\log\!\left(e^{k(f_m) \cdot \text{ICC}_\text{ref}(m) + g(f_m)} + d\right)
    - g(f_m)}{k(f_m)}.
    \label{eq:iacc_sut}
\end{equation}
Here, $d$ serves as the distortion parameter controlling the generated signal's ICC, based on the original signal's measured $\text{ICC}_\text{ref}(m)$ by the encoder.

The high-quality configuration (Q4) applies no distortion ($d = 0$), trying to preserve the full spatial image.
Q3 sets $d = 0.99 \approx 1$\,JND, the theoretical threshold of reliable detectability; in practice, expert listeners perceived this as a clear ASW reduction with lightly impaired spatial image stability.
Q2 ($d$ = Collapse) represents the ICC distortion condition in which all per-band ICC values are forced directly to unity ($\text{ICC}_\text{gen} = 1$),
causing the stereo image to collapse, except for possible ILD-generated widening.
Q1 ($d$ = Overflow) defines the lowest quality condition by setting $d$ to a negative value outside the valid range of the psychometric function, forcing the codec into an overflow state that yields the widest possible ASW with maximum decorrelation across all frequency bands\cite{delgado2019investigations}.
Expert listening during level tuning revealed that this mode produces temporal instabilities and incoherence in the reconstructed spatial image, typically rated worse than simply narrowing ASW.

In the subjective experiments, the reference is the BCC ``no distortion'' reconstruction (decoded with unmodified metadata) rather than the original audio, ensuring that perceived quality degradation comes solely from the deliberate manipulation of inter-channel cues.

\vspace{-0.5em}

\subsection{Dialogue Enhancement Processing}\label{subsec:processing2} \vspace{-0.5em}
For the \ac{DE} processing method,
a set of synthetic and real conditions are used jointly to:
i) assess spatial degradation in the DE use case via controlled mask manipulations,
ii) evaluate real-world dialogue separation systems in spatially challenging scenes, and
iii) span the MUSHRA quality scale through these complementary methods.
The channels of a stereo mixture signal are defined at a time $n$ as $ y_i(n) = s_i(n) + v_i(n)$ with channel index~$ i \in \{1, 2\}$,
where $s_i(n)$ and $v_i(n)$ denote the foreground (FG) and background (BG) signal, respectively.
In the listening test, the reference signal is $y_i(n)$, obtained from the ground-truth FG and BG.
We define the \ac{DE} task in two steps. First, given the mixture signal, the FG and BG signals are estimated as $\hat{s}_i(n)$, and $\hat{v}_i(n) = y_i(n) - \hat{s}_i(n)$.
The channels of the enhanced mixture signal are then defined as
\begin{equation}
    y_{\textrm{DE}, i}(n) = \hat{s}_i(n) + \alpha \; \hat{v}_i(n); \quad i \in \{1, 2\},
\end{equation}
where $\alpha$ denotes the attenuation factor ($16$~dB in this work).
The quality levels differ by FG estimation method, as follows.

\textbf{\Ac{qPSM}}: The FG signal is obtained by applying the \acs{PSM}\cite{erdogan2015phase} bounded to the range $[-1, 1]$. The masks are estimated in the STFT domain using the ground-truth signals and quantized using a uniform midtread quantizer (covering the range $[-1, 1]$) with $3$~bits. The masks are estimated individually for the left and right channels.

\textbf{\Acl{qms} PSM (qms\acs{PSM}$_{a, b}$)}: Ground-truth signals are used to estimate mid- and side-channel \acs{PSM} masks.
The same quantizer from \acs{qPSM} quantizes the mid-channel mask using $b$~bits and the side-channel mask using $a$~bits. Unlike {\ac{qPSM}}, the quantized masks are corrupted by additive, uniformly distributed noise in the range $[-\Delta, \Delta]$ ($\Delta$ being the quantization step size), introducing random deviations around the ideal quantized mask values.
This perturbation controls spectral and spatial degradation via mid and side channel quantization resolution, simulating suboptimal mask estimation for real-world DE distortions.

\textbf{DeepFilterNet} \cite{schroter2022deepfilternet} and \textbf{Bandit-v2} \cite{watcharasupat2024} applied to the mid/side channels separately: The pre-trained \acp{DNN}, provided by the authors, are used.

All conditions (from both Sec.~\ref{subsec:processing1}-\ref{subsec:processing2}) were loudness-normalized to $-23$~LUFS \cite{BS1770}.

\vspace{-1em}

\subsection{Listening Test}\label{subsubsec:listening} \vspace{-0.5em}

To obtain subjective ratings for the binaural cue distortion experiment (i.e., ICC and ILD) and the DE experiment, we conducted two separate MUSHRA tests.
In the binaural cue distortion experiment, 23 subjects (mean age \SI{21.0}{} years, standard deviation \SI{1.3}{}) participated, with an average of \SI{2.7}{} years of professional audio experience in areas such as mixing/production and music performance.
This experiment had 14 MUSHRA trials, each with seven conditions: four quality levels, a \SI{3.5}{\kilo\hertz} low-pass anchor, a mono anchor, and a reference.
In the DE experiment, 12 subjects (mean age \SI{21.3}{} years, standard deviation \SI{1.6}{}) participated, with an average of \SI{3.3}{} years of professional audio experience. The experiment had eight MUSHRA trials, each with nine conditions: six quality levels, a \SI{3.5}{\kilo\hertz} low-pass anchor, a mono anchor, and a reference.
All tests were performed using Beyerdynamic DT770 Pro \SI{250}{\ohm} closed-back headphones connected to a professional audio interface and a computer, running the same listening test software\footnote{\url{https://github.com/Netflix-Skunkworks/listening-test-app}} as in \cite{Torcoli2024ODAQ}.

Before the test, subjects completed six training sessions with critical material (non-overlapping with the test material) to sensitize them to the distortions. During the test, they were instructed to set a comfortable volume and avoid further adjustments. All sessions took place in acoustically treated listening rooms. Subjects were undergraduate students from the Music Media Production program who are trained musicians and received technical ear training. All subjects were attested normal hearing through an audiological test.

\begin{figure}[t]
    \centering
    \includegraphics[width=0.95\columnwidth]{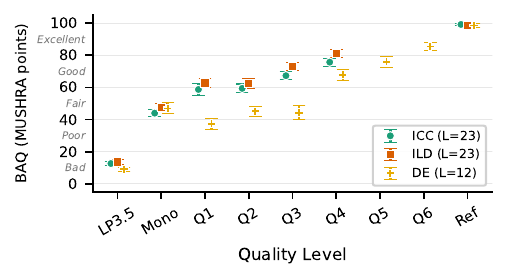}
    \vspace{-1.5em}
    \caption{Mean BAQ (in MUSHRA points) with 95\% confidence intervals across quality levels, reference, and anchor conditions. Scores are averaged over all items.
    } \vspace{-1em}
    \label{fig:mushra_proposed}
\end{figure}

\begin{figure}[t]
    \centering
    \includegraphics[width=0.9\columnwidth]{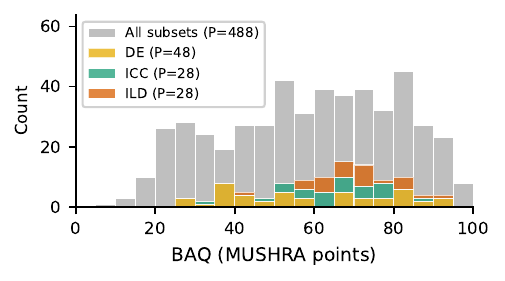}
    \vspace{-1.5em}
    \caption{Histogram of mean BAQ scores for individual items in all subsets (excluding references and anchors), with the proposed addition highlighted.}
    \label{fig:histogram}
    \vspace{-1.5em}
\end{figure}

The results of the listening tests are summarized in \Figure{fig:mushra_proposed}, which shows mean Basic Audio Quality (BAQ) and 95\% t-confidence intervals for each quality level and processing method. \Figure{fig:mushra_proposed} depicts averages over all items, showing DE results covering the
``poor'' to ``excellent'' MUSHRA range.
The distribution of scores for the ICC and ILD is narrower and mostly within the ``good'' range when averaging over items.
When individual items are inspected, ICC scores extend from ``poor'' to ``excellent'' and ILD scores from ``fair'' to ``excellent'', as indicated in \Figure{fig:histogram}.
The mono anchor corresponding to the complete loss of spatial image is rated in the ``fair'' range. This is in line with previous results for mono anchors in ODAQ \cite{Dick2025investigating} and findings in the literature \cite{rumsey2005relative} that spatial quality affects only a portion of overall quality.
Consequently, the minimum quality that can be achieved by introducing spatial distortions was expected to be limited.

\begin{table}[t]
\centering
\footnotesize
\caption{Overview of the ODAQ-Collection. Notation: (M) = MUSHRA in the laboratory with expert listeners; (C) = Crowdsourced variation of MUSHRA with self-reported experts; (P) = Audio pair count; (L) = Number of listeners; (U) = Number of unique audio sources; (PM) = Number of processing methods.}
\begin{tabularx}{\columnwidth}{@{} l c l l c X @{}}
\toprule
\textbf{Subset} & \textbf{Test} &\textbf{P (w/o anchors)} &  \textbf{L} &  \textbf{U} & \textbf{PM} \\
\midrule
2024-icassp\cite{Torcoli2024ODAQ}& M & 210 (150)   & 26    & 22  & 6    \\ \midrule

2024-aes\cite{Dick2024ODAQBSU}& M & None     & 16           & 0 & 0\\  \midrule

2025-aes\cite{Dick2025investigating}&  M   & 154 (102)   & 16          & 3 & 6\\ \midrule

2025-lanzend.\cite{lanzendorfer2025evaluating}& C & 176 (132) & 11-17  & 0  & 2\\  \midrule

\textbf{proposed addition}& M & 148 (104)   & 12-23 & 18 & 3\\
\midrule
Total:& & 688 (488) &  & 43 & 17 \\
\bottomrule
\end{tabularx}
\vspace{-1em}
\label{tab:overview}
\end{table}

\vspace{-0.5em}

\section{Overview of the Collection}\label{sec:overview} \vspace{-0.5em}

To support future expansions, we organized all subsets, including the proposed addition, into a unified resource with scripts for benchmarking and aggregating results. We refer to this resource as the ODAQ-Collection, which currently comprises five subsets and a total number of 688 processed and unprocessed audio signal pairs spanning 17 processing methods, as summarized in \Table{tab:overview}.
The collection also includes 18,984 raw subjective ratings from individual listeners, making further investigation possible.
\Figure{fig:histogram} shows the histogram of all mean BAQ scores across the entire collection, excluding references and anchors.
Overall, the distribution spans the 20 to 90 MUSHRA range relatively evenly, with modest concentration around 50 and 80 points.
Extending the benchmarks in \cite{Dick2024ODAQBSU, delgado2025exploring}, we include an in-depth benchmark\footnote{\url{https://fraunhofer-iis.github.io/ODAQ/}} of various metrics against the full collection as an evolving component of this resource.
All data\footnote{\url{https://doi.org/10.5281/zenodo.23425817}} and code\footnote{\url{https://github.com/Fraunhofer-IIS/ODAQ}} are available online.

\vspace{-1.0em}

\section{Conclusion}\label{sec:conclusion} \vspace{-0.5em}
This paper introduced a new addition to ODAQ featuring high-quality recordings of speech and music with diverse stereo images, items processed under controlled ICC, ILD, and DE distortions, and corresponding lab-based MUSHRA results.
We consolidated this addition with existing subsets into ODAQ-Collection, comprising 688 processed/unprocessed audio signal pairs spanning 17 processing method classes, with mean BAQ scores fairly uniformly distributed across the scale.
The limitations of our addition include recordings made predominantly in a single venue, a small DE listener pool, geographically concentrated listening sessions, and narrow listener demographics.
Future additions could address these limitations with recordings from additional venues and larger, more diverse listener pools.
We leave an in-depth investigation of spatial audio quality metrics, such as BINAQUAL~\cite{panah2025binaqual}, to future work,
and we hope the ODAQ-Collection provides a foundation for audio quality assessment research and invites contributions that broaden its scope.

\bibliographystyle{IEEEbib}
{\small
\bibliography{ref}
}

\end{document}